\documentclass[superscriptaddress,twocolumn,prd,aps,showpacs,secnumarabic]{revtex4}
\usepackage{graphicx}
\begin{document}

\title{Photons, neutrinos, and optical activity}
\author{Ali Abbasabadi}
\affiliation{Department of Physical Sciences,
Ferris State University, Big Rapids, Michigan 49307, USA}
\author{Wayne W. Repko}
\affiliation{Department of Physics and Astronomy,
Michigan State University, East Lansing, Michigan 48824, USA}

\date{\today}
\begin{abstract}
\vspace*{0.5cm}
\hspace*{0.1cm}
We compute the one-loop helicity amplitudes for low-energy
$\nu\gamma\to\nu\gamma$ scattering and its crossed channels in the standard
model with massless neutrinos. In the center of mass, with $\sqrt{s} =
2\omega\ll 2m_e$, the cross sections for these $2\to 2$ channels grow roughly
as $\omega^6$. The scattered photons in the elastic channel are circularly
polarized and the net value of the polarization is non-zero. We also present a
discussion of the optical activity of a sea of neutrinos and estimate the
values of its index of refraction and rotary power.
\end{abstract}
\maketitle

\vskip1pc

\section{Introduction}
\label{sec:1}

Neutrino-photon elastic scattering $\nu\gamma\to\nu\gamma$ and its crossed
channels, which are of interest in astrophysical applications, have been
studied using a variety of models for the weak interaction
\cite{cm,mjl,ls,cy,dr93,liu}. When the center of mass energy $\sqrt{s}$ is much
less than twice the electron mass $2m_e$, the amplitudes for these processes
with massless neutrinos are of leading order $s^2/m_W^4$, where $m_W$ is
$W$-boson mass \cite{dr93}. This dependence leads to low-energy cross sections
which grow as $s^3$, with a scale set by $m_W$. Apart from a factor
$\ln(m_W^2/m_e^2)$, there is no dependence on $m_e$ in leading order. Here, we
present expressions for the low-energy neutrino-photon helicity amplitudes
which are valid to higher orders in $s$, and contain terms whose scale is set
by $m_e^2$. We find that the helicity flip amplitudes for the
$\nu\gamma\to\nu\gamma$ and the helicity non-flip amplitudes for the
$\gamma\gamma\to\nu\bar\nu$, which vanish in leading order, are non-zero.
Furthermore, the inclusion of higher powers of $s$, such as $s^3/m_W^6$,
enables us to use the forward elastic scattering amplitudes for
$\nu\gamma\to\nu\gamma$ to study the optical activity of a sea of neutrinos
\cite{kn}.

In the next section, we use invariant decompositions of the amplitudes for
$\nu\gamma\to\nu\gamma$ and its crossed channels to obtain properties of and
restrictions on the corresponding helicity amplitudes. Section III gives the
numerical results for the complete one-loop helicity dependent differential
and total cross sections. This is followed by a discussion and conclusions,
which include a treatment of the production of circularly polarized photons in
low energy $\nu\gamma\to\nu\gamma$ scattering. We also calculate the index of
refraction and the rotary power for a sea of neutrinos as a function of the
energy of an incident photon and the temperature of the neutrinos.

\section{Helicity amplitudes}
\label{sec:2}

The general expressions for the Lorentz-invariant, Bose symmetric, and time
reversal invariant helicity amplitudes, ${{\cal
A}^{\nu\gamma\to\nu\gamma}_{\lambda_1\lambda_2}}(s,t,u)$, for the process
$\nu\gamma\to\nu\gamma$ can be found in Refs.\,\cite{adr01,nieves83}. They are
\begin{eqnarray} \label{ampgngn}
{{\cal A}^{\nu\gamma\to\nu\gamma}_{++}}(s,t,u) & = &
su\cos(\theta/2){\cal F}(s,t,u)\,,\label{ampgngn1} \\\
{\cal A}^{\nu\gamma\to\nu\gamma}_{--}(s,t,u) & = &
-s^2\cos(\theta/2){\cal F}(u,t,s)\,,\label{ampgngn2} \\\
{\cal A}^{\nu\gamma\to\nu\gamma}_{+-}(s,t,u) & = &
st\cos(\theta/2)[{\cal G}(s,t,u) - {\cal G}(u,t,s)]
\,,\;\;\;\label{ampgngn3} \\\
{\cal A}^{\nu\gamma\to\nu\gamma}_{-+}(s,t,u) & =  &
{\cal A}^{\nu\gamma\to\nu\gamma}_{+-}(s,t,u)\;,\label{ampgngn4}
\end{eqnarray}
where the Mandelstam variables $s$, $t$, and $u$ are defined by $s = (p_1 +
k_1)^2$, $t = (p_1 - p_2)^2 = -\frac{1}{2}s(1 - z)$, $u = (p_1 - k_2)^2 =
-\frac{1}{2}s(1 + z)$, and $z = \cos\theta$, with  $\theta$  the angle between
the incoming neutrino, which is moving in the $+z$ direction, and the outgoing
neutrino. The helicity of the incoming photon is $\lambda_1 = \pm 1$ and the
helicity of the outgoing photon is $\lambda_2 = \pm 1$. Here, the 4-momenta of
the incoming neutrino and photon are $p_1$ and $k_1$, respectively, with $p_2$
and $k_2$ denoting the corresponding outgoing momenta. To ensure the
conservation for the angular momentum in
Eqs.\,(\ref{ampgngn1})--(\ref{ampgngn3}), it is necessary to require that the
functions ${\cal F}(s,t,u)$, ${\cal F}(u,t,s)$, and $[{\cal G}(s,t,u) - {\cal
G}(u,t,s)]$ be non-singular in the limit $u \to 0$ (backward scattering). In
addition, the function $[{\cal G}(s,t,u) - {\cal G}(u,t,s)]$ must also be
non-singular in the limit $t \to 0$ (forward scattering).

From the Eqs.\,(\ref{ampgngn1})--(\ref{ampgngn4}), the interchange of $s$ and
$u$ results in the following relation
\begin{eqnarray} \label{symmetrygngn}
{\cal A}^{\nu\gamma\to\nu\gamma}_{\lambda_1\lambda_2}(s,t,u) & = &
{\cal A}^{\nu\gamma\to\nu\gamma}_{{-\lambda_1}{-\lambda_2}}(u,t,s)\,,
\end{eqnarray}
where, under this interchange, we have changed the factor
$s\cos(\theta/2) = s\sqrt{\displaystyle {-u/s}}$ to
$u\sqrt{\displaystyle {-s/u}} =
-s\sqrt{\displaystyle {-u/s}} = -s\cos(\theta/2)$.

A similar decomposition of the helicity amplitudes ${{\cal
A}^{\gamma\gamma\to\nu\bar\nu}_{\lambda_1\lambda_2}}(s,t,u)$ for the crossed
channel process $\gamma\gamma\to\nu\bar\nu$, can be obtained from
Refs.\,\cite{adr01} and \cite{addr99}. The resulting expressions are
\begin{eqnarray} \label{ampggnn}
{{\cal A}^{\gamma\gamma\to\nu\bar\nu}_{-+}}(s,t,u) & = &
\frac{1}{2}su\sin\theta{\cal F}(t,s,u)\,,\label{ampggnn1} \\\
{\cal A}^{\gamma\gamma\to\nu\bar\nu}_{+-}(s,t,u) & = &
-\frac{1}{2}st\sin\theta{\cal F}(u,s,t)\,,\label{ampggnn2} \\\
{\cal A}^{\gamma\gamma\to\nu\bar\nu}_{--}(s,t,u) & = &
\frac{1}{2}s^2\sin\theta[{\cal G}(t,s,u) - {\cal G}(u,s,t)]
\,,\;\label{ampggnn3} \\\
{\cal A}^{\gamma\gamma\to\nu\bar\nu}_{++}(s,t,u) & =  &
{\cal A}^{\gamma\gamma\to\nu\bar\nu}_{--}(s,t,u)\;,\label{ampggnn4}
\end{eqnarray}
with $s$, $t$, and $u$ defined by $s = (k_1 + k_2)^2$, $t = (k_1 - p_1)^2 =
-\frac{1}{2}s(1 - z)$, $u = (k_1 - p_2)^2 = -\frac{1}{2}s(1 + z)$, and $z =
\cos\theta$, where  $\theta$ is the angle between the incoming photon 1, which
is moving in the $+z$ direction, and the outgoing neutrino. The helicities of
the incoming photons are $\lambda_1 = \pm 1$ and $\lambda_2 = \pm 1$, and the
incoming photons have 4-momenta $k_1$ and $k_2$, while $p_1$ and $p_2$ are the
momenta of the outgoing neutrino and anti-neutrino, respectively. In this
case, conservation of angular momentum in
Eqs.\,(\ref{ampggnn1})--(\ref{ampggnn3}) requires that the functions ${\cal
F}(t,s,u)$, ${\cal F}(u,s,t)$, and $[{\cal G}(t,s,u) - {\cal G}(u,s,t)]$ be
non-singular in the limit $u \to 0$. This is sufficient to make these functions
non-singular in the limit $t \to 0$.

Using Eqs.\,(\ref{ampggnn1})--(\ref{ampggnn4}), the interchange of $t$ and $u$
results in the following relation
\begin{eqnarray} \label{symmetryggnn}
{\cal A}^{\gamma\gamma\to\nu\bar\nu}_{\lambda_1\lambda_2}(s,t,u) & = &
-{\cal A}^{\gamma\gamma\to\nu\bar\nu}_{{-\lambda_1}{-\lambda_2}}(s,u,t)\,.
\end{eqnarray}
In addition, the comparison of Eqs.\,(\ref{ampgngn1})--(\ref{ampgngn4}) with
Eqs.\,(\ref{ampggnn1})--(\ref{ampggnn4}) shows that
\begin{eqnarray} \label{symmetryggnngngn}
{\cal A}^{\gamma\gamma\to\nu\bar\nu}_{\lambda_1\lambda_2}(t,s,u) & = & -i{\cal
A}^{\nu\gamma\to\nu\gamma}_{{-\lambda_1}{\lambda_2}}(s,t,u)\,.
\end{eqnarray}
Here, under the interchange of $s$ and $t$, we have changed the factor
$s\sin\theta = 2s\sqrt{\displaystyle {tu/s^2}}$ to $2t\sqrt{\displaystyle
{su/t^2}} = -2\sqrt{\displaystyle {su}} = -2is\sqrt{\displaystyle {(1 + z)/2}}
= -2is\cos(\theta/2)$.

Using the invariance of the helicity amplitudes under the CPT operation, we
obtain the helicity amplitudes for the processes
$\bar\nu\gamma\to\bar\nu\gamma$ and $\nu\bar\nu\to\gamma\gamma$ from those of
$\nu\gamma\to\nu\gamma$ and $\gamma\gamma\to\nu\bar\nu$, respectively. The
results are
\begin{eqnarray}
{{\cal A}^{\bar\nu\gamma\to\bar\nu\gamma}_
{\lambda_1 \lambda_2}}(s,t,u) & = &
{{\cal A}^{\nu\gamma\to\nu\gamma}_
{-\lambda_1 -\lambda_2}}(s,t,u)
\,,\label{cpt1}\\\
{\cal A}^{\nu\bar\nu\to\gamma\gamma}_
{\lambda_1\lambda_2}(s,t,u) & = &
{\cal A}^{\gamma\gamma\to\nu\bar\nu}_
{-\lambda_1 -\lambda_2}(s,t,u)
\;.\label{cpt2}
\end{eqnarray}

\begin{figure}[h]
\centering\includegraphics[width=2.5in,clip]{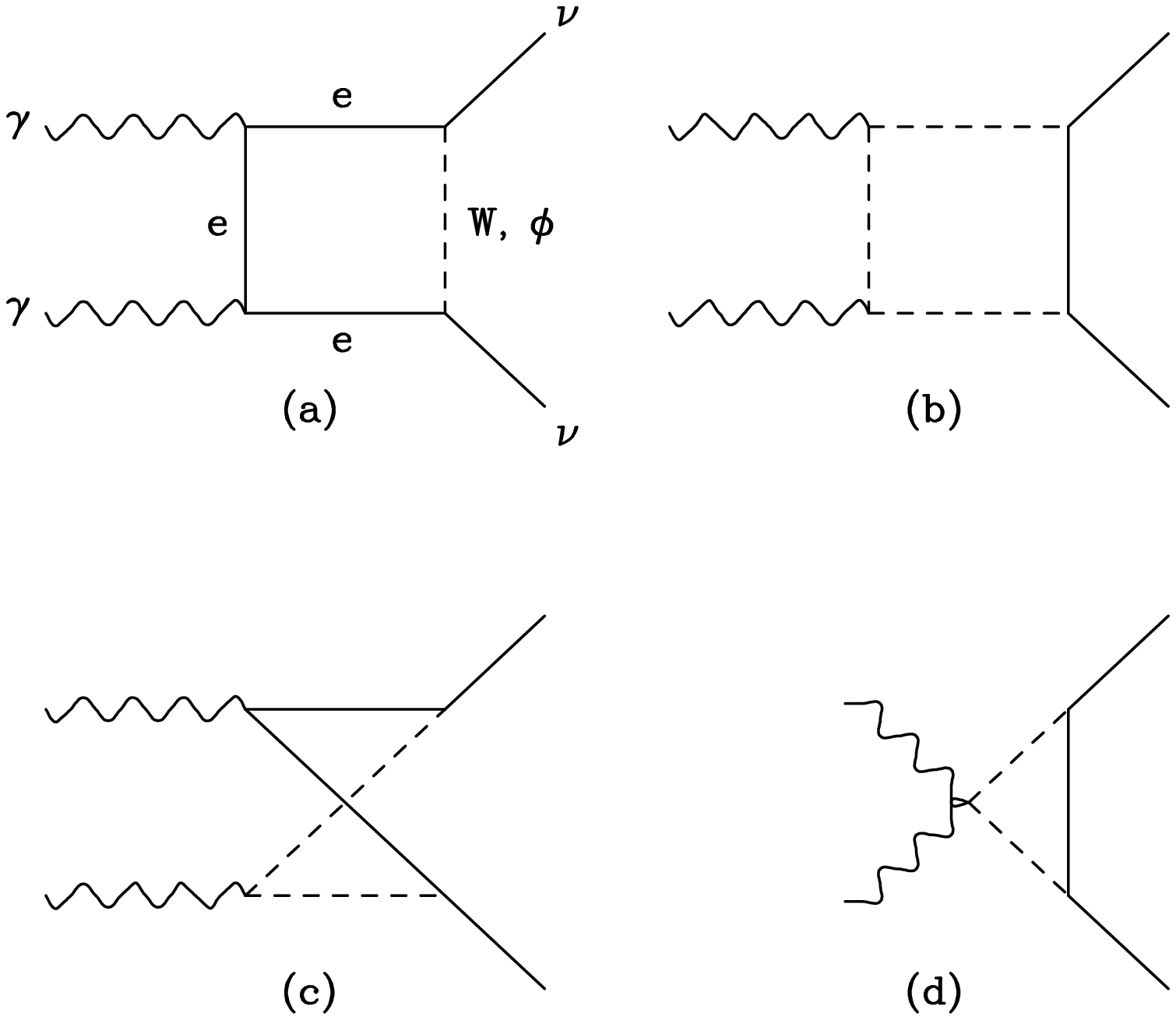}
\vspace{.10in}
\caption{Diagrams for $\nu_e\gamma \protect\to \nu_e\gamma$ or
$\gamma\gamma \protect\to \nu_e\bar\nu_e$.
The diagram (d) will give zero contribution. For each of (a), (b), (c)
there is also a diagram with the photons interchanged.} \label{diag}
\end{figure}

We have calculated the amplitudes for the diagrams of Fig.\,\ref{diag}, for
$\nu\gamma\to\nu\gamma$ and its crossed channels, in a nonlinear $R_{\xi}$
gauge such that the coupling between the photon, the $W$-boson, and the
Goldstone boson ($\phi$) vanishes \cite{dr93,gauge,pall}. Since the Goldstone
boson-electron couplings introduce a factor $m_e^2/m_W^2$, and we are keeping
terms of this order in our amplitudes, the contribution from the Goldstone
boson, in the diagrams of Fig.\,\ref{diag}, must be included. For zero
neutrino mass, the two sets of $W$-exchange and $\phi$-exchange diagrams are
separately gauge invariant. Also, the contributions of these two sets of
diagrams to the helicity amplitudes for $\nu\gamma\to\nu\gamma$ separately
have the structure of Eqs.\,(\ref{ampgngn1})--(\ref{ampgngn4}). This is also
true for the cross channel processes.

Using the algebraic manipulation software {\sc{form}} \cite{form} and
{\sc{schoonschip}} \cite{schip}, we have expressed the diagrams in terms of
Feynman parameter integrals, and for $\sqrt{s} \ll 2m_e$, have expanded these
amplitudes in a power series in $s/m_e^2$, $t/m_e^2$, and $m_e^2/m_W^2$. The
results of the calculation for the functions ${\cal F}(s,t,u)$ and ${\cal
G}(s,t,u) - {\cal G}(u,t,s)$, are
\begin{eqnarray} \label{ffgg}
{\cal F}(s,t,u) & = & \frac{\alpha^2}{8m_W^4\sin^2\theta_W}\,
f(s,t,u)\,,\label{ff1}\\
{\cal G}(s,t,u) - {\cal G}(u,t,s) & = &
\frac{\alpha^2}{8m_W^4\sin^2\theta_W}\,g(s,t,u)\;,\label{gg1}
\end{eqnarray}
where
\begin{eqnarray} \label{fstugstu}
f(s,t,u) & = & -4 -\frac{16}{3}\ln\left(\frac{m_W^2}{m_e^2}\right)
-\frac{22}{45}\frac{t}{m_e^2}\nonumber\\
&&
-\frac{11}{315}\frac{t^2}{m_e^4}
-\frac{37}{9450}\frac{t^3}{m_e^6}
-\frac{4}{7425}\frac{t^4}{m_e^8} \nonumber\\
&&
+\frac{m_e^2}{m_W^2}\left[\left( \frac{8}{3} -\frac{8}{3}\frac{s}{m_e^2}
+\frac{8}{3}\frac{t}{m_e^2}\right)
\ln\left(\frac{m_W^2}{m_e^2}\right)\right.\nonumber\\
&&
-10 +\frac{64}{9}\frac{s}{m_e^2} -\frac{163}{45}\frac{t}{m_e^2}
 -\frac{4}{45}\frac{st}{m_e^4}\nonumber\\
&&
+\frac{169}{630}\frac{t^2}{m_e^4} -\frac{1}{315}\frac{st^2}{m_e^6}
+\frac{71}{3780}\frac{t^3}{m_e^6}
\nonumber\\
&&\left.
-\frac{1}{4725}\frac{st^3}{m_e^8}
+\frac{106}{51975}\frac{t^4}{m_e^8}\right]
,\label{fstu1}\\ [4pt]
g(s,t,u) & = & \frac{s-u}{m_e^2}\left[\frac{2}{45}
+\frac{1}{315}\frac{t}{m_e^2}+\frac{1}{3150}\frac{t^2}{m_e^4}\right.
\nonumber\\
&&
+\frac{2}{51975}\frac{t^3}{m_e^6}
+\frac{m_e^2}{m_W^2}\left(-\frac{17}{45} -\frac{23}{630}\frac{t}{m_e^2}
\right.\nonumber\\
&&\left.\left.
-\frac{1}{420}\frac{t^2}{m_e^4}
-\frac{23}{103950}\frac{t^3}{m_e^6} \right)\right].\label{gstu1}
\end{eqnarray}
Here, $\theta_W$ is the weak mixing angle, $\alpha$ is the fine structure
constant, and we have neglected higher powers of $s/m_e^2$, $t/m_e^2$, and
$m_e^2/m_W^2$. The first two terms of $f(s,t,u)$ in Eq.\,(\ref{fstu1}) were
previously derived in the Ref.\,\cite{dr93}. These results show that there are
many higher order terms whose scale is set by $m_e^2$. Note, however, that in
the forward direction ($t=0$) the scale in Eq.\,(\ref{fstu1}) is set by
$m_W^2$. This suggests that Eq.\,(\ref{fstu1}) is valid even for $\sqrt{s} >
m_e$ when $t=0$.

We can confirm the validity of the forward limit of Eq.\,(\ref{fstu1}) for the
range $\sqrt{s}\ll m_W$ by using the dispersion relation
\begin{eqnarray}\label{disp1}
{{\cal A}^{\nu\gamma\to\nu\gamma}_{\lambda\lambda}}(s,0,-s) & = &
\frac{s^2}{\pi}\int_{(m_W+m_e)^2}^{\infty}\frac{ds'}{s'}
\nonumber\\
&&
\times\left(\frac{\sigma_{\lambda}(s')}{s'-s}
+\frac{\sigma_{\lambda}(-s')}{s'+s}\right),
\end{eqnarray}
to obtain the exact value of the non-flip forward helicity amplitude ${{\cal
A}^{\nu\gamma\to\nu\gamma}_{\lambda\lambda}}(s,0,-s)$ for $s < (m_W + m_e)^2$.
The cross section ${\sigma}_{\lambda} \equiv \sigma^{\nu\gamma\to
e^-W^+}_{\lambda}$ is the total cross section for the process $\nu\gamma\to
e^-W^+$, after summation over the helicities of the $W$-boson and the electron
($\lambda$ is helicity of the photon). Using Eqs.\,(\ref{symmetrygngn}) and
(\ref{disp1}), it can be shown that the following symmetry relation must exist
\begin{equation} \label{symmetrysigngew}
{\sigma}_{\lambda}(-s') = {\sigma}_{-\lambda}(s')\,.
\end{equation}
Therefore, we can write \cite{adr01}
\begin{eqnarray}\label{disp2}
{{\cal A}^{\nu\gamma\to\nu\gamma}_{\lambda\lambda}}(s,0,-s) & = &
\frac{s^2}{\pi}\int_{(m_W+m_e)^2}^{\infty}\frac{ds'}{s'}
\nonumber\\
&&
\times\left(\frac{\sigma_{\lambda}(s')}{s'-s}
+\frac{\sigma_{-\lambda}(s')}{s'+s}\right).
\end{eqnarray}
A calculation of $\sigma_{\lambda}(s)$ gives
\begin{eqnarray}\label{signgew}
\sigma_{\lambda}(s) & = &
\sqrt{2}G_F\alpha\left[
\left(\frac{2}{s} -\lambda\frac{m_e^2}{s^2}
+ 6\lambda\frac{m_W^2}{s^2}
- 2\frac{m_e^2m_W^2}{s^3}\right.\right. \nonumber \\
& &\left.\left.  - 2\frac{m_e^4}{s^3}  + 4\frac{m_W^4}{s^3}\right)
\sqrt{\mu(s,m_e^2,m_W^2)}\right. \nonumber \\
& &\left.
+\left(- 2\lambda\frac{m_W^2}{s} + \frac{m_e^2m_W^2}{s^2}
+ 3\lambda\frac{m_e^2m_W^2}{s^2}
+ \frac{m_e^4}{s^2}\right.\right. \nonumber \\
& &\left.\left.
- 2\frac{m_W^4}{s^2} + 2\lambda\frac{m_W^4}{s^2} +
\frac{m_e^2m_W^4}{s^3}
- 2\frac{m_e^4m_W^2}{s^3}\right.\right. \nonumber \\
& &\left.\left.
 - \frac{m_e^6}{s^3} + 2\frac{m_W^6}{s^3}\right)
\ell(s,m_e^2,m_W^2)\right. \nonumber \\
& &\left.
+\left(\frac{1}{2}\frac{m_e^2}{s} + \frac{\lambda}{2}\frac{m_e^2}{s}
+ \frac{m_W^2}{s} -\lambda\frac{m_W^2}{s}
\right.\right. \nonumber \\
& &\left.\left.
+ \frac{m_e^2m_W^2}{s^2} + 3\lambda\frac{m_e^2m_W^2}{s^2}
- 2\frac{m_W^4}{s^2} + 2\lambda\frac{m_W^4}{s^2}
\right.\right. \nonumber \\
& &\left.\left.
+ \frac{m_e^4}{s^2} + \frac{m_e^2m_W^4}{s^3}
- 2\frac{m_e^4m_W^2}{s^3}\right.\right. \nonumber \\
& &\left.\left.
- \frac{m_e^6}{s^3} + 2\frac{m_W^6}{s^3}\right)
\ell(s,m_W^2,m_e^2)\right],
\end{eqnarray}
where
\begin{eqnarray}
\mu(x,y,z) & = &x^2+y^2+z^2-2xy-2xz-2yz\,, \\
\ell(x,y,z) & =
&\ln\left(\frac{x-y+z+\sqrt{\mu(x,y,z})}{x-y+z-\sqrt{\mu(x,y,z})}
\right)\,,
\end{eqnarray}
and $G_F = \pi\alpha/(\sqrt{2}\,m_W^2\sin^2{\theta_W})$ is the Fermi coupling.
After setting the electron mass to zero everywhere but in the logarithm, our
spin averaged cross section $(\sigma_+ + \sigma_-)/2$ agrees with the result
previously obtained by Seckel \cite{seck}.

Assuming $\sqrt{s} \ll m_W$, an expansion of the dispersion integral
Eq.\,(\ref{disp2}) to order $s^3/m_W^6$ gives
\begin{eqnarray}
{{\cal A}^{\nu\gamma\to\nu\gamma}_{++}}(s,0,-s) & = &
\frac{-\alpha^2s^2}{8m_W^4\sin^2\theta_W}f(s,0,-s)\,,\label{disp3}\\\
{{\cal A}^{\nu\gamma\to\nu\gamma}_{--}}(s,0,-s) & = &
\frac{-\alpha^2s^2}{8m_W^4\sin^2\theta_W}f(-s,0,s)\,,\label{disp4}
\end{eqnarray}
where
\begin{eqnarray} \label{fstu2}
f(s,0,-s) & = & -4 -\frac{16}{3}\ln\left(\frac{m_W^2}{m_e^2}\right)
\nonumber\\
&&
+\frac{m_e^2}{m_W^2}\left[\left( \frac{8}{3}
-\frac{8}{3}\frac{s}{m_e^2} \right)
\ln\left(\frac{m_W^2}{m_e^2}\right)\right.\nonumber\\
&&
\left.
-10 +\frac{64}{9}\frac{s}{m_e^2}\right]\,,
\end{eqnarray}
and we have neglected terms in higher powers of $m_e^2/m_W^2$. The function
$f(-s,0,s)$ can be found from Eq.\,(\ref{fstu2}) by replacing $s$ by $-s$.
Notice that Eqs.\,(\ref{disp3})--(\ref{fstu2}) are consistent with
Eqs.\,(\ref{symmetrygngn}) and (\ref{fstu1}), and although Eq.\,(\ref{fstu1})
was derived for $\sqrt{s} \ll 2m_e$, its range of validity for $t = 0$, as
suggested above, extends to $\sqrt{s} \ll m_W$.

\section{Differential and total cross sections}

\begin{figure}[h]
\centering\includegraphics[width=3.00in,clip]{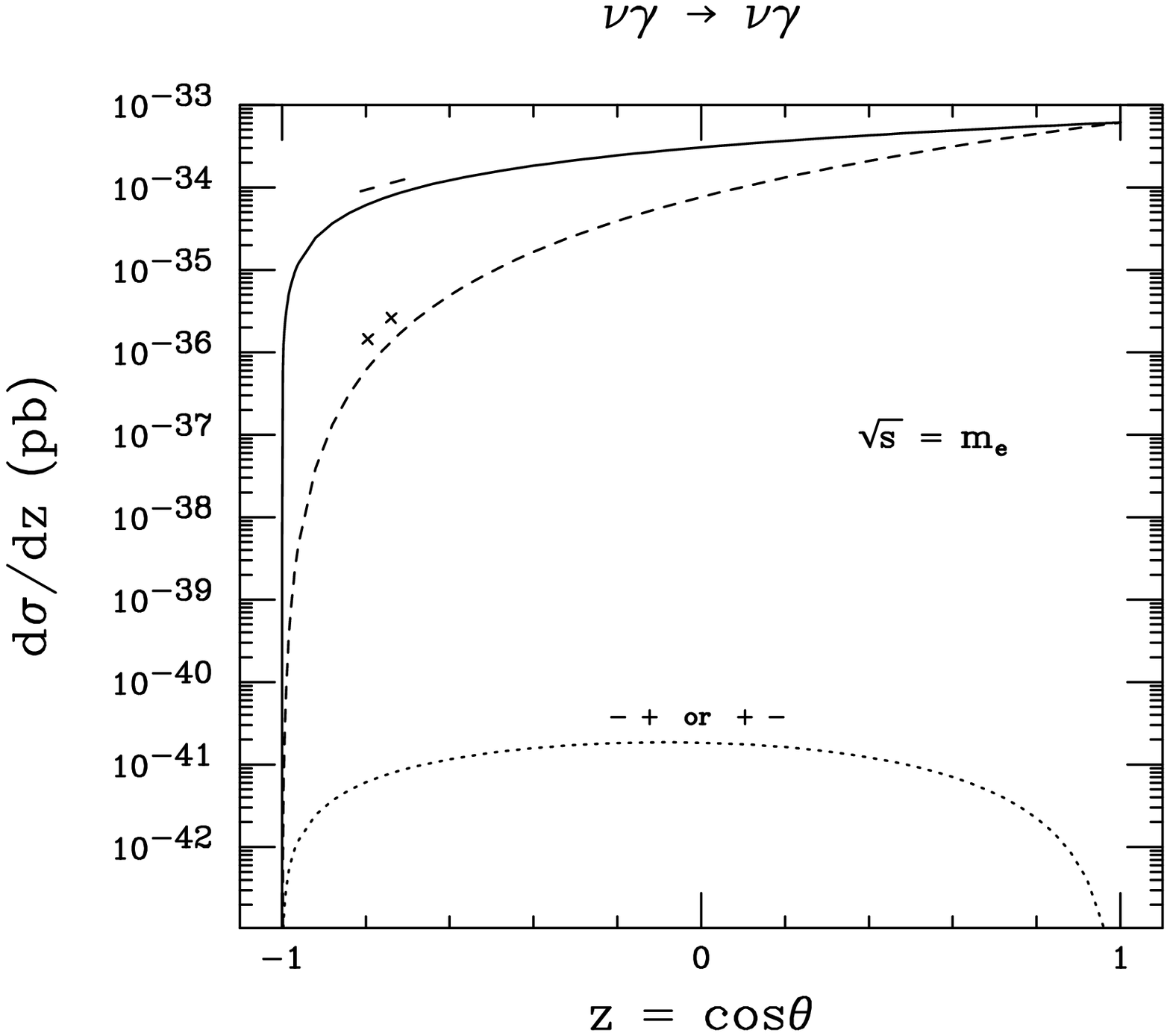}
\vspace{.10in}
\caption{The helicity dependent differential cross sections for
$\nu\gamma\protect\to\nu\gamma$ are shown for
$\protect\sqrt{s} = m_e$. The solid line is $d\sigma_{--}/dz$,
the dashed line is $d\sigma_{++}/dz$, and the dotted line is
$d\sigma_{+-}/dz$. The $d\sigma_{-+}/dz$ is the same as
$d\sigma_{+-}/dz$.} \label{gngn_dsigma}
\end{figure}

In Fig.\,\ref{gngn_dsigma}, we show the differential cross sections, for
$\nu\gamma\to\nu\gamma$, using
\begin{equation}\label{dsiggngn}
\frac{d{\sigma}^{\nu\gamma\to\nu\gamma}_{\lambda_1\lambda_2}}{dz}
= \frac{1}{32\pi s}|
{\cal A}^{\nu\gamma\to\nu\gamma}_{\lambda_1\lambda_2}|^2\,,
\end{equation}
where $\lambda_1$ and $\lambda_2$ are the helicities of the incoming and
outgoing photons, respectively. Here, $z = \cos\theta$, and  $\theta$ is the
angle between the incoming neutrino, which is moving in the $+z$ direction,
and the outgoing neutrino. Figure\,\ref{gngn_dsigma} shows the identity of the
two helicity amplitudes ${\cal A}^{\nu\gamma\to\nu\gamma}_{+-}$ and ${\cal
A}^{\nu\gamma\to\nu\gamma}_{-+}$ given by Eq.\,(\ref{ampgngn4}). It also shows
the vanishing of the amplitudes for backward scattering, and vanishing of the
flip amplitudes for forward scattering, as given in
Eqs.\,(\ref{ampgngn1})--(\ref{ampgngn4}).

\begin{figure}[h]
\centering\includegraphics[width=3.00in,clip]{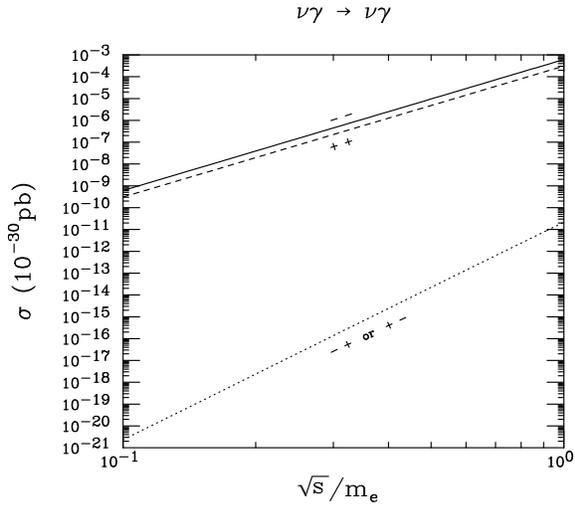}
\vspace{.10in}
\caption{The helicity dependent total cross sections for
$\nu\gamma\protect\to\nu\gamma$ are shown. The solid line is
$\sigma_{--}$, the dashed line is $\sigma_{++}$, and the
dotted line is $\sigma_{+-}$. The $\sigma_{-+}$ is the
same as $\sigma_{+-}$.} \label{gngn_sigma}
\end{figure}

The total cross sections for helicities $\lambda_1$ and $\lambda_2$ are given
by
\begin{equation}\label{siggngn}
\sigma^{\nu\gamma\to\nu\gamma}_{\lambda_1\lambda_2} =
\int_{-1}^1
\frac{d{\sigma}^{\nu\gamma\to\nu\gamma}_{\lambda_1\lambda_2}}
{dz}\,dz\,,
\end{equation}
and are plotted in Fig.\,\ref{gngn_sigma}. Also shown in dots is the helicity
flip cross section, which can be seen to be much smaller than the cross
sections for helicity non-flip. This feature seems not to be a consequence of
any symmetry. Figure\,\ref{gngn_sigma} illustrates the roughly $s^3$ behavior
of the total cross section for helicity non-flip, and $s^5$ behavior for the
helicity flip, at photon energies $\omega \ll m_e$. A fit to the points in
Fig.\,\ref{gngn_sigma} gives
\begin{eqnarray}\label{siggngnlow}
\sigma^{\nu\gamma\to\nu\gamma}_{--} & = & 3.9\times 10^{-32}
\left(\frac{\omega}{m_e}\right)^6\,\mbox{\rm pb}\,,\label{siggngnlowmm}\\\
\sigma^{\nu\gamma\to\nu\gamma}_{++} & = & 2.0\times 10^{-32}
\left(\frac{\omega}{m_e}\right)^6\,\mbox{\rm pb}\,,\label{siggngnlowpp}\\\
\sigma^{\nu\gamma\to\nu\gamma}_{-+} & = & 2.2\times 10^{-38}
\left(\frac{\omega}{m_e}\right)^{10}\,\mbox{\rm pb}\,,\label{siggngnlowmp}
\end{eqnarray}
with
\begin{equation}
\sigma^{\nu\gamma\to\nu\gamma}_{+-}  = \sigma^{\nu\gamma\to\nu\gamma}_{-+}\,.
\label{siggngnlowpmmp}
\end{equation}
Here, $\omega = \sqrt{s}/2$ is the energy of a photon (or a neutrino), and
$m_e$ is the mass of the electron. Therefore, the total cross section for an
unpolarized initial photon can be approximated as
\begin{equation}\label{unpolarsiggngnlow}
\sigma^{\nu\gamma\to\nu\gamma} = 3.0\times 10^{-32}
\left(\frac{\omega}{m_e}\right)^6\,\mbox{\rm pb}\,,\;\;\;\;
\omega\ll m_e\,.
\end{equation}
This is the same $\omega^6$ behavior as found in the numerical calculations of
Ref.\,\cite{adr01} for the region $m_e \ll \omega \ll m_W$, but with a larger
slope.

\begin{figure}[h!]
\centering\includegraphics[width=3.00in,clip]{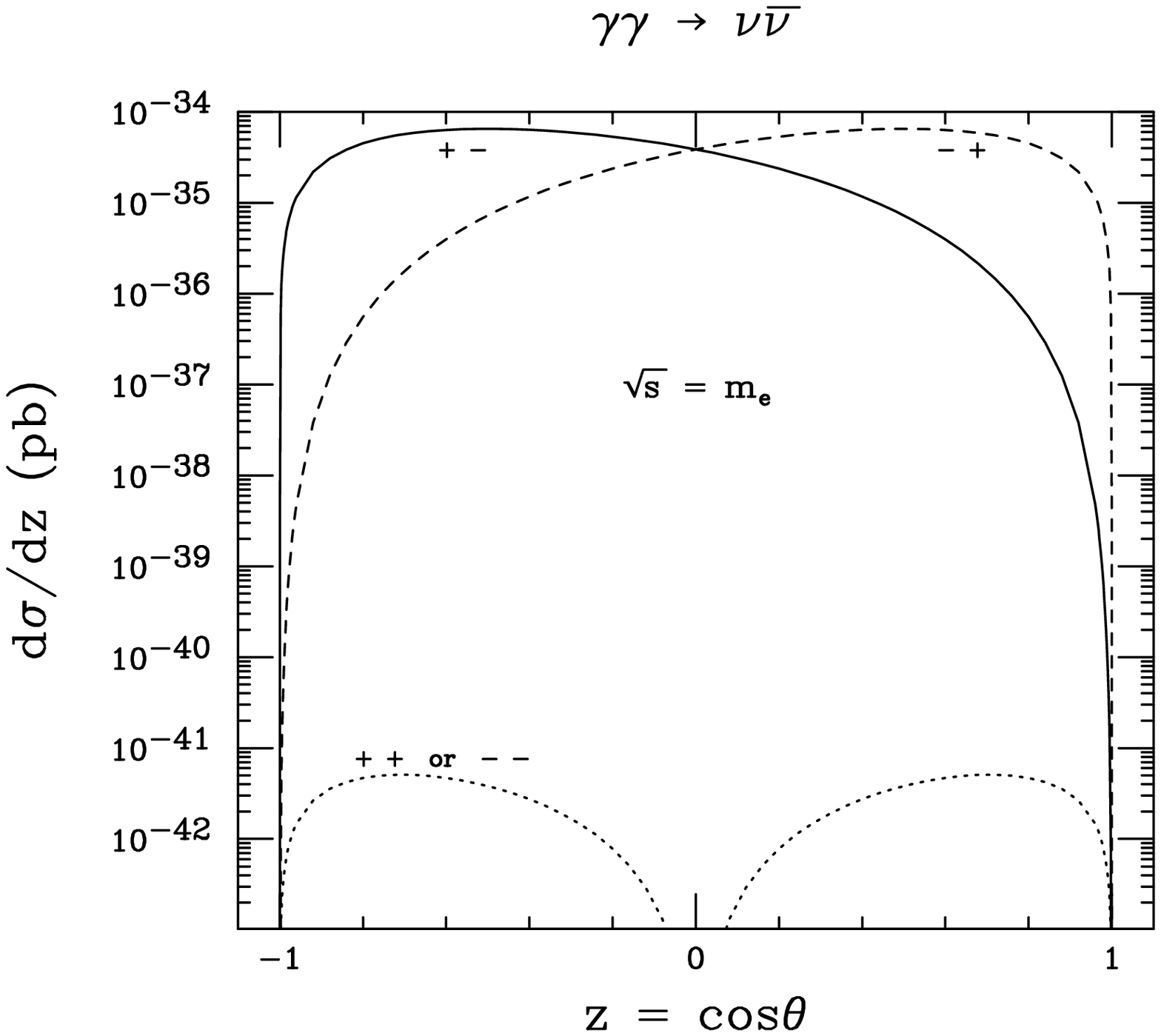}
\vspace{.10in}
\caption{The helicity dependent differential cross sections for
$\gamma\gamma\protect\to\nu\bar\nu$ are shown for
$\protect\sqrt{s} = m_e$. The solid line is $d\sigma_{+-}/dz$,
the dashed line is $d\sigma_{-+}/dz$, and the dotted line is
$d\sigma_{++}/dz$. The $d\sigma_{--}/dz$ is the same as
$d\sigma_{++}/dz$.} \label{ggnn_dsigma}
\end{figure}

In Fig.\,\ref{ggnn_dsigma}, we show the differential cross sections for
$\gamma\gamma\to\nu\bar\nu$, using
\begin{equation}\label{dsigggnn}
\frac{d{\sigma}^{\gamma\gamma\to\nu\bar\nu}_{\lambda_1\lambda_2}}{dz}
= \frac{1}{32\pi s}|
{\cal A}^{\gamma\gamma\to\nu\bar\nu}_{\lambda_1\lambda_2}|^2\,,
\end{equation}
where $\lambda_1$ and $\lambda_2$ are the helicities of the incoming photons.
In this case, $z = \cos\theta$, with $\theta$ being the angle between the
incoming photon $1$, which is moving in the $+z$ direction, and the outgoing
neutrino. The identity of the two helicity amplitudes ${\cal
A}^{\gamma\gamma\to\nu\bar\nu}_{++}$ and ${\cal
A}^{\gamma\gamma\to\nu\bar\nu}_{--}$, as implied by Eq.\,(\ref{ampggnn4}), is
also shown in this figure, as is the vanishing of the amplitudes for the
forward and backward scattering, implied by
Eqs.\,(\ref{ampggnn1})--(\ref{ampggnn4}). Notice also that this figure clearly
exhibits the symmetry relations of Eq.\,(\ref{symmetryggnn}).

\begin{figure}[h!]
\centering\includegraphics[width=3.00in,clip]{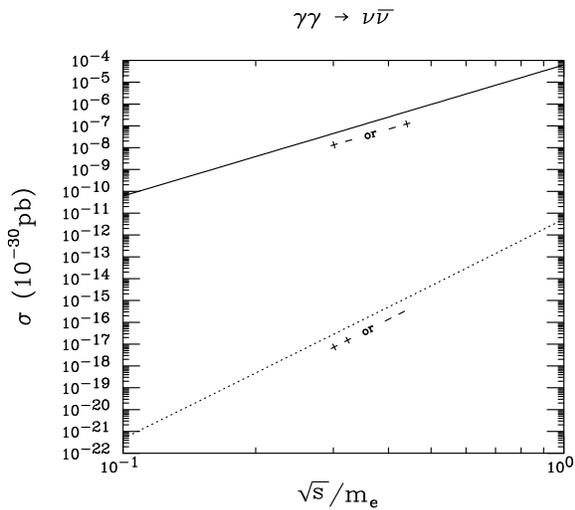}
\vspace{.10in}
\caption{The helicity dependent total cross sections for
$\gamma\gamma\protect\to\nu\bar\nu$ are shown. The solid line is
$\sigma_{+-}$ and the dotted line is $\sigma_{++}$. The
$\sigma_{-+}$ is the same as $\sigma_{+-}$, and
$\sigma_{--}$ is the same as $\sigma_{++}$.} \label{ggnn_sigma}
\end{figure}

The total cross sections for $\gamma\gamma\to\nu\bar\nu$ are plotted in
Fig.\,\ref{ggnn_sigma}, using
\begin{equation}\label{sigggnn}
\sigma^{\gamma\gamma\to\nu\bar\nu}_{\lambda_1\lambda_2} = \int_{-1}^1
\frac{d{\sigma}^{\gamma\gamma\to\nu\bar\nu}_{\lambda_1\lambda_2}} {dz}\,dz\,.
\end{equation}
Shown in dots is the cross section for the helicity non-flip, which can be
seen to be much smaller than the helicity flip cross sections. Again, we see a
roughly $s^3$ behavior of the dominant contributions to the total cross
section, and a $s^5$ behavior of the non-leading contributions, at photon
energies $\omega \ll m_e$. Fits to the points in this case give
\begin{eqnarray}\label{sigggnnlow}
\sigma^{\gamma\gamma\to\nu\bar\nu}_{--} & = & 4.0\times 10^{-33}
\left(\frac{\omega}{m_e}\right)^6\,\mbox{\rm pb}
\,,\label{sigggnnlowmm}\\\
\sigma^{\gamma\gamma\to\nu\bar\nu}_{-+} & = & 5.5\times 10^{-39}
\left(\frac{\omega}{m_e}\right)^{10}\,\mbox{\rm pb}
\,,\label{sigggnnlowmp}
\end{eqnarray}
with
\begin{eqnarray}
\sigma^{\gamma\gamma\to\nu\bar\nu}_{++} & = &
 \sigma^{\gamma\gamma\to\nu\bar\nu}_{--}
\,,\label{sigggnnlowppmm}\\\
\sigma^{\gamma\gamma\to\nu\bar\nu}_{+-} & = &
\sigma^{\gamma\gamma\to\nu\bar\nu}_{-+}
\,,\label{sigggnnlowpmmp}
\end{eqnarray}
which are the consequences of Eq.\,(\ref{symmetryggnn}). The total cross
section for the unpolarized initial photons can be approximated as
\begin{equation}\label{unpolarsigggnnlow}
\sigma^{\gamma\gamma\to\nu\bar\nu} = 2.0\times 10^{-33}
\left(\frac{\omega}{m_e}\right)^6\,\mbox{\rm pb}\,,\;\;\;\;
\omega\ll m_e\,.
\end{equation}
In this case, too, the $\omega^6$ dependence is the same as that found in
Ref.\,\cite{addr99} for $m_e\ll\omega\ll m_W$, but with a large slope.

The differential cross sections for the process $\bar\nu\nu\to\gamma\gamma$,
for photons with helicities $\lambda_1$ and $\lambda_2$, can be obtained from
the following relation
\begin{equation}
\frac{d{\sigma}^{\bar\nu\nu\to\gamma\gamma}_{\lambda_1\lambda_2}}{dz} =
\frac{d{\sigma}^{\gamma\gamma\to\nu\bar\nu}_{-\lambda_1-\lambda_2}}{dz}
\label{dsignngg}\,,
\end{equation}
where $z = \cos\theta$, and on the left side of this equation, $\theta$ is the
angle between the incoming anti-neutrino, which is moving in the $+z$
direction, and the outgoing photon 1 with helicity $\lambda_1$. The various
helicity-dependent total cross sections are related as
\begin{eqnarray}
\sigma^{\bar\nu\nu\to\gamma\gamma}_{\lambda\lambda} &=&
\frac{1}{2}\sigma^{\gamma\gamma\to\nu\bar\nu}_{-\lambda-\lambda}\,, \\
\sigma^{\bar\nu\nu\to\gamma\gamma}_{\lambda-\lambda} &=&
\sigma^{\gamma\gamma\to\nu\bar\nu}_{-\lambda\lambda}\,,
\end{eqnarray}
and total cross section for the production of a pair of back-to-back photons
can be obtained from
\begin{equation}
{\sigma}^{\bar\nu\nu\to\gamma\gamma} = \frac{1}{2!}\,\sum_{\lambda_1\lambda_2}
\int_{-1}^{+1}
\frac{d{\sigma}^{\bar\nu\nu\to\gamma\gamma}_{\lambda_1\lambda_2}} {dz}\,dz
\label{signnngg3}\,.
\end{equation}
In view of Eq.\,(\ref{dsignngg}), we have
\begin{equation}
{\sigma}^{\bar\nu\nu\to\gamma\gamma}
=
2\,{\sigma}^{\gamma\gamma\to\nu\bar\nu}
\label{signnngg4}\,.
\end{equation}

\section{Discussion and conclusions}

We have shown that the energy dependence of the cross sections
$\sigma^{\nu\gamma\to\nu\gamma}$ and $\sigma^{\gamma\gamma\to\nu\bar\nu}$ at
low energies, $\omega\ll m_e$, is the same as that in the energy region, $m_e
\ll \omega \ll m_W$. As a result, the effective interaction introduced in
Refs.\,\cite{adr01} and \cite{dkr} contains all the essential of the features
of these cross sections. This includes the prediction that the final photons
in the channel $\nu\gamma\to\nu\gamma$ acquire (parity violating) circular
polarization.

To investigate the degree of circular polarization of the final photon in the
process $\nu\gamma\to\nu\gamma$, we define the polarization P as
\begin{equation}\label{p1}
\rm P = \frac{\sigma_{--} + \sigma_{+-} - \sigma_{-+} - \sigma_{++}}
{\sigma_{--} + \sigma_{+-} + \sigma_{-+} + \sigma_{++}}\,,
\end{equation}
where ${\sigma}_{\lambda_1\lambda_2} \equiv
\sigma^{\nu\gamma\to\nu\gamma}_{\lambda_1\lambda_2}$ is defined in
Eq.\,(\ref{siggngn}). It is clear from
Eqs.\,(\ref{siggngnlowmm})--(\ref{siggngnlowpmmp}) that $\sigma_{+-} =
\sigma_{-+} \ll \sigma_{--}$, and $\sigma_{--}\simeq 2\,\sigma_{++}$.
Therefore, for photons with the energies $\omega \ll m_e$, Eq.\,(\ref{p1})
gives
\begin{equation}\label{p2}
\rm P \simeq \frac{1}{3}\,,
\end{equation}
which is independent of the $\omega$. This result in comparable to that found
in  Ref.\,\cite{adr01}, where it was shown that P is about 0.3 for photons
with energies $1$ GeV $\lesssim \omega\lesssim 10$ GeV.

\begin{figure}[h]
\centering\includegraphics[width=3.00in,clip]{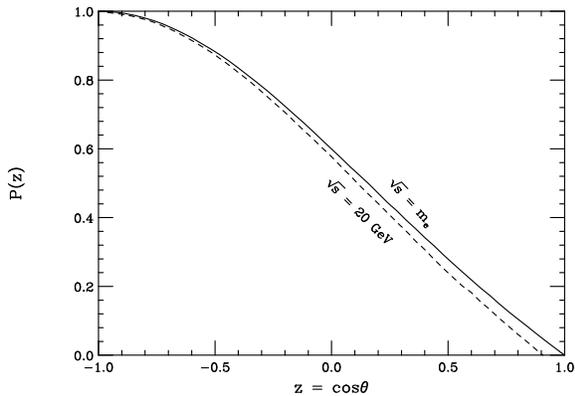}
\vspace{.10in}
\caption{The polarization $\rm{P}(z)$ of the final photons in the process
$\nu\gamma\to\nu\gamma$, as it is defined in the Eq.\,(\ref{p4}), is shown.
The solid line is polarization for the center of mass energy $\sqrt{s} = m_e$,
while the dashed line is for the center of mass energy $\sqrt{s} = 20$ GeV,
which is taken from the Ref.\,\cite{adr01}.} \label{polarization_angle}
\end{figure}

The angular dependence, ${\rm P}(z)$, of the final photon's polarization in the
process $\nu\gamma\to\nu\gamma$, can be obtained from the differential form of
Eq.\,(\ref{p1}),
\begin{equation}\label{p4}
{\rm P}(z) = \frac{d\sigma_{--} /dz - d\sigma_{++} /dz} {d\sigma_{--} /dz +
2d\sigma_{+-} /dz +d\sigma_{++} /dz}\,\,.
\end{equation}
where the $d{\sigma}_{\lambda_1\lambda_2}/dz \equiv
d\sigma^{\nu\gamma\to\nu\gamma}_{\lambda_1\lambda_2}/dz$ are defined in the
Eq.\,(\ref{dsiggngn}), and we have used the equality $d\sigma_{+-}/dz =
d\sigma_{-+}/dz$. The polarization ${\rm P}(z)$ is plotted in
Fig.\,\ref{polarization_angle} as a function of $z = \cos\theta$, where
$\theta$ is the angle between the incoming and the outgoing neutrinos. In this
figure, the solid line is for photons of energy $\omega = m_e/2$, and the
dashed line, which is taken from the Ref.\,{\cite{adr01}}, is for photons of
energy $\omega = 10$ GeV . It is clear from Fig. \,\ref{polarization_angle}
that the polarization ${\rm P}(z)$ remains effectively unchanged for wide
range of energies $\omega \lesssim 10$ GeV. Notice that Eq.\,(\ref{p4}) can be
approximated as
\begin{equation}\label{p5}
{\rm P}(z) \simeq \frac{4-(1+z)^2}{4+(1+z)^2}\,\,,
\end{equation}
which is independent of the energy of photon.

Our low-energy helicity non-flip amplitudes, obtained from
Eqs.\,(\ref{ampgngn1}) and (\ref{ampgngn2}) using Eq.\,(\ref{fstu1}), enable us
to discuss the optical activity of a sea of neutrinos \cite{kn}. To do this, it
is necessary to establish a relationship between the forward-scattering
amplitude and the Lorentz-invariant helicity amplitude for the general case
where photons and neutrinos are colliding non-collinearly. Following M{\o}ller
\cite{m,r}, we define the Lorentz-invariant differential cross section
$d\sigma$, for the general non-collinear process $1+2 \to 3+4$ as \cite{n}
\begin{equation}\label{dsig1}
d{\sigma} = \frac{|{\cal A}(s,t,u)|^2}{F}\,dQ\,,
\end{equation}
where ${\cal A}(s,t,u)$ is the Lorentz-invariant amplitude, $F$ is the
Lorentz-invariant flux
\begin{equation}\label{flux}
F = 4\,\sqrt{(p_1\cdot p_2)^2 - m^2_1\,m^2_2}\,,
\end{equation}
and $dQ$ is the Lorentz-invariant phase space differential element
\begin{eqnarray}\label{dq1}
dQ &=& (2\pi)^4 \delta^{(4)} (p_1+p_2-p_3-p_4)\nonumber\\
&& \times \frac{d^3\vec{p}_3}{(2\pi)^3\,2E_3}
\frac{d^3\vec{p}_4}{(2\pi)^3\,2E_4}\,.
\end{eqnarray}
After integration over $d^3\vec{p}_4$, we have
\begin{equation}\label{dq2}
dQ = \frac{{\vec{p}_3}^2}{16\pi^2 E_3 E_4}
{d\Omega_3 \over\displaystyle
{\left| \frac{|\vec{p}_3|}{E_3}
- \frac{{\vec{p}_3}.{\vec{p}_4}}{|\vec{p}_3| E_4}\right|}}\,.
\end{equation}
Here, $m_i$, $E_i$, and $\vec{p}_i$ are the mass, energy, and momentum of the
particle $i$, respectively $(i=1,2,3,4)$. The resulting $d\sigma$ is
essentially identical to  Eq. (93) of Ref. \cite{r}. From Eqs.\,(\ref{dsig1}),
(\ref{flux}), and (\ref{dq2}), it is clear that, in the case of forward
elastic scattering ($m_3 = m_1, m_4 = m_2; \vec{p}_3 = \vec{p}_1, \vec{p}_4 =
\vec{p}_2$) and for massless particles ($m_i = 0, i=1,2,3,4$), we have
\begin{equation}\label{dsig2}
\left.\frac{d\sigma}{d\Omega_3}\right|_{t=0} = \frac{E^2_1|{\cal
A}(s,0,-s)|^2}{16\pi^2 s^2}\,,
\end{equation}
where $s = (p_1 +p_2)^2 = 4E_1 E_2\,{\sin}^2(\theta_{12}/2)$, and
$\theta_{12}$ is the angle between the momenta $\vec{p}_1$ and $\vec{p}_2$ of
the initial incoming particles $1$ and $2$. A comparison of Eq.\,(\ref{dsig2})
with the cusomary definition of the scattering amplitude, written for forward
scattering as
\begin{equation}\label{dsig3}
\left.\frac{d\sigma}{d\Omega_3}\right|_{t=0} =  |f(0)|^2\,,
\end{equation}
gives
\begin{equation}\label{fscat1}
f(0) = \frac{E_1}{4\pi s}
\,{\cal A}(s,0,-s)\,\,.
\end{equation}

To compute the optical activity of a neutrino sea, we consider a photon of
helicity $\lambda$ and energy $\omega$ traversing a bath of neutrinos that are
in thermal equilibrium at the temperature $T_\nu$. To give an order of
magnitude estimate of the index of refraction $n_{\lambda}$ of this sea, we
write \cite{ll,gw}
\begin{eqnarray}\label{n1}
n_{\lambda} - 1 &=& \frac{2\pi}{\omega^2}
\int dN_{\nu}\,f_{\lambda\lambda}^{\nu\gamma\to\nu\gamma}(0)
\,,
\end{eqnarray}
where the forward-scattering amplitude
$f_{\lambda\lambda}^{\nu\gamma\to\nu\gamma}(0)$, from the Eq.\,(\ref{fscat1}),
is
\begin{eqnarray}
f_{\lambda\lambda}^{\nu\gamma\to\nu\gamma}(0) &=&
 \frac{\omega}{4\pi s}
{\cal A}^{\nu\gamma\to\nu\gamma}_{\lambda\lambda}(s,0,-s) \,.\label{fscat}
\end{eqnarray}
The Fermi-Dirac distribution, $dN_\nu$, is
\begin{eqnarray}
dN_{\nu} &=& \frac{1}{(2\pi)^3}
\frac{d^{3}\vec{p}_\nu}{e^{E_{\nu}/T_\nu} + 1}
\,,\label{dn}
\end{eqnarray}
and we have neglected the chemical potential for the neutrinos \cite{p93}.
Here, ${\cal A}^{\nu\gamma\to\nu\gamma}_{\lambda\lambda}(s,0,-s)$ is given in
the Eqs.\,(\ref{disp3})--(\ref{fstu2}), $\vec{p}_\nu$ and $E_\nu$ are the
momentum and energy of a neutrino, and $s = 4\omega E_\nu\,
{\sin}^2(\theta_{\nu\gamma}/2)$, where $\theta_{\nu\gamma}$ is the angle
between the incoming photon and the incoming neutrino. Therefore,
Eqs.\,(\ref{n1}) and (\ref{fscat}) give
\begin{eqnarray}\label{n2}
n_{\lambda} - 1 &=&
\int \frac{dN_{\nu}}{2\omega s}\,
{\cal A}^{\nu\gamma\to\nu\gamma}_{\lambda\lambda}(s,0,-s)
\,\,.
\end{eqnarray}

After using Eq.\,(\ref{dn}) for $dN_\nu$, Eqs.\,(\ref{disp3})--(\ref{fstu2})
for the amplitudes ${\cal A}^{\nu\gamma\to\nu\gamma}_{\lambda\lambda}(s,0,-s)$,
and performing integration in the Eq.\,(\ref{n2}), we obtain
\begin{eqnarray}
n_{+} - 1 &=& \frac{T_\nu^4}{m_W^4}\,c_0
+\frac{\omega T_\nu^5}{m_W^6}\,c_1\,,\label{np1}\\\
n_{-} - 1 &=& \frac{T_\nu^4}{m_W^4}\,c_0
-\frac{\omega T_\nu^5}{m_W^6}\,c_1\,,\label{nm1}
\end{eqnarray}
where (for $\alpha=1/137$)
\begin{eqnarray}
c_0  &=& \frac{7\alpha^2 \zeta (4)}{4\pi^2 \sin^2\theta_W}
\left[\ln\left(\frac{m_W^2}{m_e^2}\right) + \frac{3}{4}\right]
\simeq 1.1 \times 10^{-3},\;\;\;\;\;\label{c1}\\\
c_1  &=& \frac{15\alpha^2 \zeta (5)}{4\pi^2 \sin^2\theta_W}
\left[\ln\left(\frac{m_W^2}{m_e^2}\right) - \frac{8}{3}\right] \simeq 2.0
\times 10^{-3},\label{c2}
\end{eqnarray}
and $\zeta(x)$ is the Riemann zeta function. The range of the validity of the
above relations for the index of refraction, as far as energy is concerned, is
related to that of Eqs.\,(\ref{disp3})--(\ref{fstu2}), which is $s = 4\omega
E_\nu\, {\sin}^2(\theta_{\nu\gamma}/2)
 \ll m_W^2$. In Eq.\,(\ref{n2}), if we change the
upper limit of the integration on $E_\nu$ from the infinity to $fT_\nu$, the
contributions of this integral to $c_0$ and $c_1$ in the Eqs.\,(\ref{np1}) and
(\ref{nm1}) change by 9\% and 18\%, respectively, if we use $f = 7$ (for $f =
8$, the corresponding changes are 4\% and 10\%). Here, we set the following
criterion
\begin{equation}
4\omega fT_\nu  \ll \, m_W^2 \,,\label{criterion1}
\end{equation}
which for $f = 7$ is
\begin{equation}
\omega T_\nu \ll 2.7\times10^{15} \;\;{\rm GeV}\cdot{\rm K}
\,,\label{criterion2}
\end{equation}
where $\omega$ is the photon energy in GeV, and $T_\nu$ is the neutrino
temperature in Kelvin.

From Eqs.\,(\ref{np1})--(\ref{c2}), we have
\begin{eqnarray}
n_{+} - n_{-} &=&
2\frac{\omega T_\nu^5}{m_W^6}\,c_1
\nonumber\\
&\simeq & 7.0 \times 10^{-80}
\,\omega T_\nu^5
\,,\label{npm1}
\end{eqnarray}
and the following approximate relation
\begin{eqnarray}
n_{+} - 1 \simeq  n_{-} - 1 & \simeq &
\frac{T_\nu^4}{m_W^4}\,c_0
\nonumber\\
& \simeq & 1.5 \times 10^{-63}T_\nu^4
\,.\label{npm2}
\end{eqnarray}
Equation\,(\ref{npm2}) implies that the index of refraction is independent of
the helicity and the energy of the incident photon, as long as
Eq.\,(\ref{criterion2}) is satisfied.

When linearly polarized light propagates through a medium that has different
indices of refraction for positive and negative helicities ($n_+ \not= n_-$),
the plane of polarization of the light rotates by an angle $\phi$, which is
\cite{b}
\begin{equation}
\phi = \frac{\pi}{\lambda_{\gamma}}\,(n_{+} - n_{-})\,l =
\frac{\omega}{2}\,(n_{+} - n_{-})\,l
\,,\label{phi1}
\end{equation}
where $\omega$ and $\lambda_{\gamma} = 2\pi/\omega$ are the energy and
wavelength of the photon and $l$ is the distance traveled by photons in the
medium. To estimate the specific rotary power, $\phi/l$, for a sea of
neutrinos, we use Eqs.\,(\ref{npm1}) and (\ref{phi1}) to obtain
\begin{eqnarray}
\frac{\phi}{l} &=&
\frac{\omega^2 T_\nu^5}{m_W^6}\,c_1
\nonumber\\
& \simeq &
1.8 \times 10^{-64}\,\omega^2 T_\nu^5 \;\;\rm{rad/m} \,.\label{phi2}
\end{eqnarray}
A positive angle of rotation, $\phi > 0$, that is $n_+ > n_-$, corresponds to
a clockwise rotation (dextrorotation) of the plane of polarization of the
linearly polarized incident photons, as viewed by an observer that is
detecting the forward-scattered light. Thus, the optical activity of a
neutrino sea is that of a dextrorotary medium. In addition, it is clear from
Eq.\,(\ref{phi2}) that the rotary power, $\phi/l$, varies as
$1/\lambda_{\gamma}^2$, which is the same as that of quartz and most
transparent substances for visible light.

To get a rough estimate of rotation angle $\phi$ for linearly polarized photons
propagating through the relic neutrino sea, we use Eq.\,(\ref{phi2}) with $l =
ct$, $c = 3\times10^8$ m/s, $t \sim 15\times10^9$ years, $T_\nu \sim 2 $\,K,
and $\omega \sim 10^{20}$ eV, and for the neutrino part of the sea we find
\begin{equation}
\phi \sim 8\times10^{-15}\;\; {\rm rad}
\,,\label{phi3}
\end{equation}
which is exceedingly small. The antineutrino part of the sea gives
a rotation with opposite sign, such that if the asymmetry
parameter \cite{lp}, $L \equiv (N_\nu - N_{\bar\nu})/N_\gamma $,
is zero, the resultant angle of rotation $\phi$ will be zero. The
proper treatment for the case $L \not= 0$, is to include the
chemical potential in the neutrino and antineutrino distribution
functions.

\acknowledgments We wish to thank Duane Dicus for numerous helpful
discussions. One of us (A.A.) wishes to thank the Department of Physics and
Astronomy at Michigan State University for its hospitality and computer
resources. This work was supported in part by the National Science Foundation
under Grant No. PHY-0070443.

%\newpage
\section*{Figures}

\end{document}